\documentclass[prl,twocolumn,groupedaddress]{revtex4-1}
\usepackage{graphicx}
\usepackage{dcolumn}
\usepackage{bm}

\usepackage{amsmath}
\usepackage{graphicx}
\usepackage{color}
\usepackage{wrapfig}
\usepackage{epstopdf}

\begin{document}
\title{Enhancement of artificial magnetism via resonant bianisotropy}

\author{Dmitry Markovich,$^{1*}$ Kseniia Baryshnikova,$^{1*}$ Alexander Shalin,$^{1}$\\ Anton Samusev,$^{1}$ Alexander Krasnok,$^{1}$ Pavel Belov,$^{1}$ and Pavel Ginzburg$^{1,2}$}

\address{$^1$ ITMO University, St.~Petersburg 197101, Russia\\
$^2$ School of Electrical Engineering, Tel Aviv University, Ramat Aviv, Tel Aviv 69978, Israel\\
$^*$ Equal contribution}

\begin{abstract}
All-dielectric "magnetic light" nanophotonics based on high refractive index nanoparticles allows controlling magnetic component of light at nanoscale without having high dissipative losses. The artificial magnetic optical response of such nanoparticles originates from circular displacement currents excited inside those structures and strongly depends on geometry and dispersion of optical materials. Here a new approach for increasing magnetic response via resonant bianisotropy effect is proposed and analyzed. The key mechanism of enhancement is based on electric-magnetic interaction between two electrically and magnetically resonant nanoparticles of all-dielectric dimer nanoantenna. It was shown that proper geometrical arrangement of the dimer in respect to the incident illumination direction allows flexible control over all vectorial components of magnetic polarizability, tailoring the later in the dynamical range of 100$\%$ and enhancement up to 36$\%$ relative to performances of standalone spherical particles. The proposed approach provides pathways for designs of all-dielectric metamaterials and metasurfaces with strong magnetic response.
\end{abstract}

\maketitle

\section{Introduction}

Intrinsic magnetic polarizabilities of natural materials have strong frequency dependence with the fundamental cut-off in GHz range, originating from relatively low spin and orbital susceptibilities~\cite{polydoroff-high-frequency-1960}. Recently, effective polarization currents in subwavelength structured loops, organized in ordered arrays, became sources of high-frequency artificial magnetism~\cite{smith-composite-2000}. Nanostructured noble metals, supporting localized plasmon resonances could serve as building blocks for metamaterials with artificial magnetic polarizability~\cite{shalaev-optical-2007}. However, inherent material losses set severe limitations on performances of such structures~\cite{khurgin-scaling-2011}. Another approach for obtaining magnetic optical response is to employ circular displacement currents in high-index dielectric nanoparticles~\cite{kuznetsov-magnetic-2012}. This is the essence of so-called all-dielectric nanophotonics which opened the way to control magnetic component of light at nanoscale without high-dissipation, inherent for metallic nanostructures~\cite{Popa-2008, evlyukhin-demonstration-2012, Miroshnichenko-2012, ginn-realizing-2012, fu-directional-2013, Slovick-2013, Shcherbakov-2014}. The "magnetic light" concept found use in various applications, such as nanoantennas~\cite{Krasnok-2014, KrasnokOE}, quantum interface for NV-centers~\cite{KrasnokLPR}, photonic topological insulators~\cite{Slobozhanyuk2015}, broadband perfect reflectors~\cite{Krishnamurthy_13}, waveguides~\cite{Savelev2014_1}, cloacking~\cite{cloaking2015all}, harmonics generation~\cite{ShcherbakovNL2014}, wave-front engineering, and dispersion control~\cite{Staude_15}.
\begin{figure}[!b]
\centering
\includegraphics[width=0.49\textwidth]{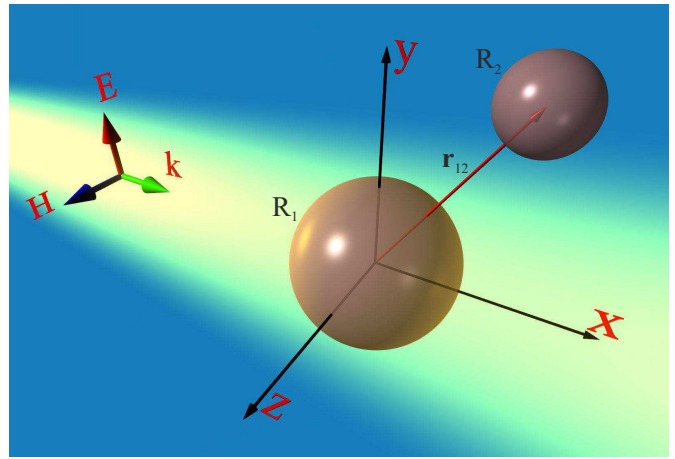}
 \caption{The geometry of the bianisotropic all-dielectric dimer nanoantenna. The nanoantenna consists of two dielectric nanoparticles separated by a distance $\mathbf {r}_{21}$. The sizes of nanoparticles are $R_1$ (bigger nanoparticle) and $R_2$ (smaller nanoparticle). The bigger nanoparticle exhibit electric dipole resonance, while the smaller one magnetic dipole resonance at the same wavelength (480 nm). The nanoantenna is excited by a $y$-polarized plane wave propagating along $x$-axis.}
 \label{fig_01:geometry}
\end{figure}

Magnetic response of a dielectric particle strongly depends on it's refractive index, shape, and external environment. The eigen frequencies of electric and magnetic resonances could span the entire visible range. However, the value of those multipole moments is limited by dispersion properties of optical materials.

Here a new approach for tailoring magnetic response of dielectric nanoparticles via resonant effect of bianisotropy is proposed. Microscopically, bianisotropy is the effect of magneto-electric coupling, where electric polarization induces magnetic and vice versa. The  signature of this effect appears in the constitutive relations, e.g the dependence of electrical induction also on magnetic field and magnetic induction also on electric field~\cite{Wegener10, Tretyakov2015}. Bianisotropy is used for achieving high values of effective polarizability in metamaterials~\cite{Marques2002}, unique properties of metasurfaces~\cite{Tretyakov2015}, and designed directivity of nanoantennas~\cite{Koenderink2013}. Previously, it was shown that the interaction of dielectric nanoparticles with substrates may increase induced magnetic moment due to the effect of non-resonant bianisotropy~\cite{markovich-magnetic-2014}. The approach proposed here is based on electric-magnetic interaction between two resonant nanoparticles of dimer nanoantenna (see Fig.~\ref{fig_01:geometry}). The nanoparticles were designed in the following way: the electric dipolar resonance of the bigger sphere overlaps with magnetic response of the smaller one. In this case, the effect of bi-resonance anisotropy is achieved: the resonant electric moment of the bigger nanoparticle induces the additional magnetic moment in the smaller one, tailoring its overall response.

The manuscript is organized as follows: first, the optical properties of isolated spherical particles are briefly discussed in the context of the resonance tuning. Next, the analytical coupled dipoles formulation of the problem is developed and verified by numerical modeling. The expression describing the magnetic moment of the nanoparticle, considering the bianisotropy, effect is derived. Furthermore, it was shown that proper geometrical arrangement of the dimer in respect to the incident illumination direction allows achieving  additional vectorial component of magnetic polarization.

\section{Coupled dipoles theory and numerical results}

\begin{figure}[!t]
\centering
\includegraphics[width=0.5\textwidth]{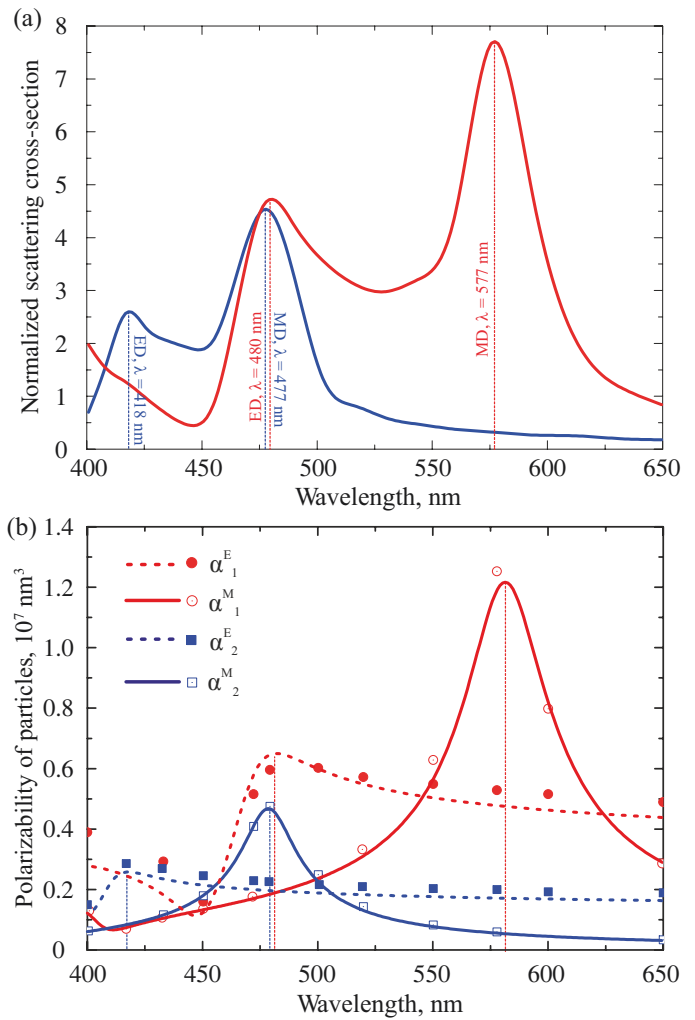}
\caption{Optical properties of single silicon nanoparticles with radii $R_1=70$ nm (red curves) and $R_2=52$ nm (blue curves). (a) The scattering cross-section spectra, normalized to geometric cross-section ($\pi r^2$). (b) Dispersion of particles polarizabilities. The curves correspond to analytical calculations, red circles and blue squares represent numerical results. Electric dipolar resonance of 70-nm radius particle overlaps with the magnetic dipolar resonance of 52-nm sphere at $\lambda\approx 480$ nm.}
\label{fig_02:standalone}
\end{figure}

In order to obtain the dimer design, properties of isolated components will be briefly discussed. First, isolated silicon sphere of radius $R_2=52$ nm, having magnetic dipolar resonance at wavelength 477 nm, i.e. in the visible range is considered. The material dispersion of crystalline silicone (c-Si) is taken from~\cite{palik-handbook-1997}. The scattering cross-section of the nanoparticle have been calculated using CST Microwave Studio, and corresponding results as the function of wavelength are presented in Fig.~\ref{fig_02:standalone}(a) (blue curve). Electric dipolar resonance is blue shifted in respect to the magnetic one and appears on the spectrum at 418 nm wavelength. This Mie resonances hierarchy is red shifted when the radius of nanoparticle is increased. It is worth noting, that high-order quadrupole resonances, also contributing to the scattering cross-section, are suppressed in this region due to perceptible losses of silicon. The bigger sphere with radius of $R_1=70$ nm exhibits its electric resonance at the similar spectral position where 52-nm sphere exhibits resonant magnetic response (Fig.~\ref{fig_02:standalone}a, red curve). The electric and magnetic polarizabilities which are associated with induced electric and magnetic moments have been also calculated (Fig.~\ref{fig_02:standalone}b). Results of numerical simulations (scatter plots) are in good agreement with analytical Mie theory (lines plot) verifying the validity of the numerical tool. Values of those electric and magnetic moments will be subsequently used in the analytical model, based on discrete dipoles approximation.

Next, the scattering of a plane wave on all-dielectric dimer nanoantenna will be analyzed. The electromagnetic scattering problem could be solved by employing Coupled Electric and Magnetic Dipole Approximation~\cite{Hopkins2015ACS} (CEMDA). In this method, complex nanostructures are represented by converging series of point electric and magnetic dipoles~\cite{novotny-principles-2012}, while the problem of two spheres could be approximated by only two. This approximation is particularly accurate, if the gap between the spheres is bigger than their radii~\cite{evlyukhin-optical-2010}. The goal of the subsequent analytical analysis is obtaining a simple formulation, underlining the interference phenomena, affecting the magnetic dipolar polarizability of the smaller particle and revealing the bianisotropic nature of the interaction. The high-order modes and their interplay have been taken into account using the full-wave numerical calculations using the CST Microwave Studio. Following the CEMDA method, total electromagnetic fields [both electric (${\bf E}$) and magnetic (${\bf H}$)] were decomposed into incident and scattered components:
\begin{eqnarray}
{\mathbf{E}(\mathbf{r}_i)}&=& {\mathbf {E}_0(\mathbf{r}_i)}+{\mathbf {E}_j^{sc}(\mathbf{r}_i)},\nonumber\\
{\mathbf{H}(\mathbf{r}_i)}&=&{\mathbf{H}_0(\mathbf{r}_i)}+{\mathbf{H}_j^{sc}(\mathbf{r}_i)},
\label{eq:e_h_general}
\end{eqnarray}
where indices $i,j=1,2~(i\neq j)$ denote first (bigger) and second (smaller) nanoparticles, respectively, $\mathbf {E}_0$ and $\mathbf{H}_0$ are electric and magnetic fields of the incident plane wave, $\mathbf {E}(\mathbf{r}_1)\equiv\mathbf {E}(0)$ and $\mathbf {H}(\mathbf{r}_1)\equiv\mathbf {H}(0)$ are the full electric and magnetic fields at the position of the first (electric) nanoparticle (with radius-vector $\mathbf {r}_1\equiv0$), $\mathbf {E}_2^{sc}(\mathbf{r}_1)$ and $\mathbf {H}_2^{sc}(\mathbf{r}_1)$ are the electric and magnetic fields scattered by the second (smaller) nanoparticle at the first particle center, and analogously for the second particle. Both particles are polarized by the incident field, as well as by the scattered one:
\begin{eqnarray}
\mathbf {p}_i=\varepsilon_0\alpha_i^E\mathbf {E}(\mathbf{r}_i),~~~\mathbf {m}_i=\alpha_i^M\mathbf {H}(\mathbf{r}_i),
\label{eq:p_m_general}
\end{eqnarray}
where $\alpha_i^E$ and $\alpha_i^M$ are the electric and magnetic polarizabilities of single particle $"i"$:
\begin{eqnarray}\label{eq2}
\alpha_i^E=j\frac{6\pi\varepsilon_{h}}{k^3_{h}}a_1, \;\;
\alpha_i^M=j\frac{6\pi}{k^3_{h}}b_1,
\end{eqnarray}
and $\varepsilon_0$ is the dielectric constant, $j$ is the imaginary unit. The coefficients $a_1$ and $b_1$ are called the first order Mie scattering amplitudes, and they can be written in simplified form which is suitable for analysis: $a_1(\lambda) = [A -B ]/[C -D ],
\; b_1(\lambda) =[B n^{-2}-A ]/[D n^{-2}-C ]$, where the following notation is introduced:
\begin{eqnarray}\label{eq4}
A&=&\left[\frac{\cos(\rho n)}{\rho n}+\sin(\rho n)-\frac{\sin(\rho n)}{(\rho n)^2}\right]\cdot\left(\frac{\sin(\rho )}{\rho^2 }-\frac{\cos(\rho )}{\rho }\right), \nonumber \\
B&=& n^2\left[\frac{\cos(\rho )}{\rho }+
\sin(\rho )-\frac{\sin(\rho )}{\rho^2 }\right]\cdot\left(\frac{\sin(\rho n)}{(\rho n)^2}
-\frac{\cos(\rho n)}{\rho n}\right), \nonumber \\
C&=& -\left[\frac{\cos(\rho n)}{\rho n}+
\sin(\rho n)-\frac{\sin(\rho n)}{(\rho n)^2}\right]\cdot\left(\frac{1}{\rho }+\frac{j}{\rho^2 }\right)e^{j\rho }, \nonumber \\
D&=& n^2\left[\frac{e^{j\rho }(j+\rho -j\rho^2 )}{\rho^2 }\right]\cdot\left(\frac{\sin(\rho n)}{(\rho n)^2}-\frac{\cos(\rho n)}{\rho n}\right),
\end{eqnarray}
where $n=\sqrt{\varepsilon/\varepsilon_h}$, $\rho =k_{h}R_i$, $R_i$ - radius of the sphere $"i"$, $\varepsilon$ - the permittivity of the material (silicon, here), $k_{h}=2\pi\sqrt{\varepsilon_h}/\lambda$ is the wave number in the surrounding space, and $\lambda$ is a wavelength.

The values of single particle polarizabilities, calculated using Eqs.~(\ref{eq2}) and (\ref{eq4}), are in a perfect agreement with our numerical calculations [see Fig.\ref{fig_02:standalone}(b)]. Thus, the scattered fields of the dipoles can be obtained through the Green function of a point dipole in free-space $\widehat{G}(\mathbf{r}_i,\mathbf{r}_j)\equiv\widehat{G}_{ij}$ ~\cite{novotny-principles-2012}, following the method CEMDA:
\begin{eqnarray}\label{eg:all_p_m}
\mathbf{p}_i &=&\alpha_i^E\left[\mathbf{E}_0(\mathbf{r}_i)+\frac{k^2_0}{\varepsilon_0}\left(\widehat{G}_{ij}\mathbf{p}_j+\frac{j}{c k_0}[\mathbf{g}_{ij}\times \mathbf{m}_j]\right)\right],\nonumber \\
\mathbf{m}_i
&=&\alpha_i^M\left[\mathbf{H}_0(\mathbf{r}_i)+k^2_0\left(\widehat{G}_{ij}\mathbf{m}_j-\frac{j c}{k_0}[\mathbf{g}_{ij}\times
\mathbf{p}_j]\right)\right],
\end{eqnarray}
where $k_0$ and $c$ are the wavenumber and speed of light in vacuum. The Green's function of a point dipole in free-space is well known:
\begin{eqnarray}\label{dopolnitelno}
& &\widehat{G}_{ij} = \frac{e^{jk_0r_{ij}}}{4\pi}\left[G^1(r_{ij})\hat{\mathbf{I}}+G^2(r_{ij})\frac{\mathbf{r}_{ij}\otimes\mathbf{r}_{ij}}{r_{ij}^2}\right], \nonumber \\
& &G^1(r_{ij}) = \left(\frac{j}{r_{ij}}+\frac{j}{k_0 r_{ij}^2}-\frac{j}{k_0^2r_{ij}^3}\right), \nonumber \\
& &G^2(r_{ij})= \left(-\frac{j}{r_{ij}}-\frac{3j}{k_0 r_{ij}^2}+\frac{3}{k_0^2r_{ij}^3}\right), \nonumber \\
& &\mathbf{g}_{ij}=\frac{e^{jk_0 r_{ij}}}{4\pi r_{ij}}\left(\frac{j k_0}{r_{ij}}-\frac{j}{r^2_{ij}}\right)\mathbf{r}_{ij},
\end{eqnarray}
where $\mathbf{r}_{ij}=\mathbf{r}_i-\mathbf{r}_{j}$ is the radius vector, connecting the center of the first dipole (coordinate origin, or center of the bigger particle) with the second one (center of the smaller particle). The tensor $\widehat{G}_{ij}$ and vector $\mathbf{g}_{ij}$ have the following symmetry of indices permutation: $\widehat{G}_{ij}=\widehat{G}_{ji}$ and $\mathbf{g}_{ij}=-\mathbf{g}_{ji}$.

Both nanoparticles, being isolated, have 3-fold degenerated (magnetic and electric) dipolar resonances, oriented along the unit vectors of a Cartesian coordinate system. The excitation of an isolated sphere is solely defined by the polarization of the incident wave -- for example, linearly polarized beam will excite only one of 3 components of the dipolar mode. However, the geometry of the coupled dipoles together with the excitation (not necessarily coinciding with one of the symmetry axis of the system) will break the degeneracy and, as the result, all three vectorial components must be taken into account. The obtained set of equations can be solved analytically by means of the matrix inversion or numerically in the same fashion. In order to verify the validity of the proposed theoretical model, we consider a particular case, where the system of Eqs.~(\ref{eg:all_p_m}) has a simple and intuitive solution. Arranging the nanoparticles along the x-axis and exciting the system with linearly polarized plane wave along y-axis with angles being $\theta=\pi/2$ and $\varphi=0$, the set of 12 coupled equations was reduced just for 4, since the symmetry considerations allow the moment components to be induced only along z-axis (magnetic dipole) and y-axis (electric dipole). Similar configuration was studied in~\cite{noskov-nonlinear-2012} for hybrid metal-dielectric nanoantennas with non-resonant dielectric nanoparticles. This set of coupled equations has particularly simple solution, which agrees well with the numerical calculations. The values of the magnetic polarizability enhancement, calculated both analytically and numerically as the function of the distance $\mathbf{r}_{12}$ between nanoparticles are shown in Fig.~\ref{fig_03:simple_m2z}. It is clearly seen, that the analytical and numerical {models} agree with each other and showing 15$\%$ enhancement of the magnetic moment of the single nanoparticle, as the result of the bianisotropic coupling. For clear understanding of oscillatory behaviour of the magnetic moments, the simplest model, when the polarizabilities $\alpha_1^M$ and $\alpha_2^E$ are supposed to be zero, was considered. In this case the simple formula can be obtained in form showing the bianisotropic nature of the effect:
\begin{figure}[!t]
\centering
\includegraphics[width=0.5\textwidth]{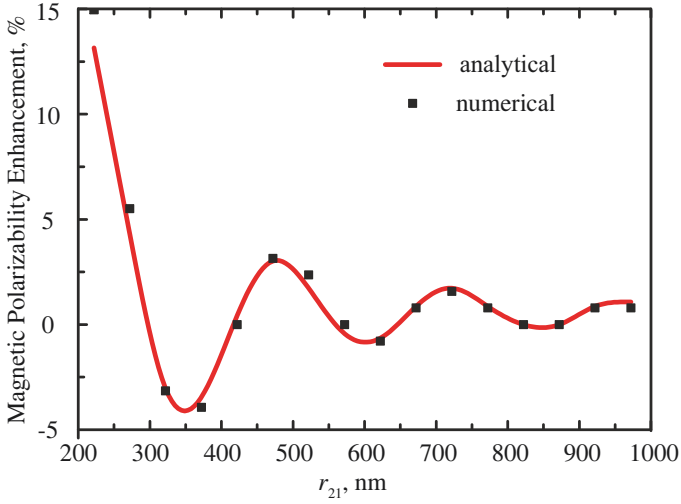}
\caption{Enhancement of magnetic polarizability of the smaller nanoparticle for the case $\theta=\pi/2$, $\varphi=0$. Red curve corresponds to the solution Eq.~(\ref{eq:simple_m2z}), black squares are the results of numerical full-wave simulation.}
 \label{fig_03:simple_m2z}
\end{figure}
\begin{eqnarray}
& &m_{2}=\eta H_{0}+ \gamma E_{0},\nonumber\\
& &\eta=\frac{\alpha_2^M e^{jk_0r_{21}}}{1-\alpha_1^E \alpha_2^M g_{ij}^2k_0^2},\nonumber\\
& &\gamma=\frac{-\varepsilon_0 \alpha_1^E \alpha_2^M g_{ij} j k_0 c}{1-\alpha_1^E \alpha_2^M g_{ij}^2k_0^2}
 \label{eq:simple_m2z}
\end{eqnarray}
The qualitative analysis of magnetic moment $m_2$ enables to observe the clear interference phenomena -- the direct excitation of magnetic dipole by the plane wave [$\alpha_2^M e^{ik_0r}$ term in Eq.~(\ref{eq:simple_m2z})] and the contribution of the scattered field through the term $\alpha_2^E \alpha_1^M$. Thus, oscillation behavior of magnetic moment enhancement is the result of the interference phenomenon (see Fig.~\ref{fig_03:simple_m2z}). For longer distances the coupling between the particles becomes weaker and converges to the value of the isolated particle. An almost perfect fit of full numerical modeling with the CEMDA method enables to use the later for analysis of more complex structures without involving full wave simulations. This approach will be subsequently employed.
\begin{figure}[!t]
\centering
\includegraphics[width=0.5\textwidth]{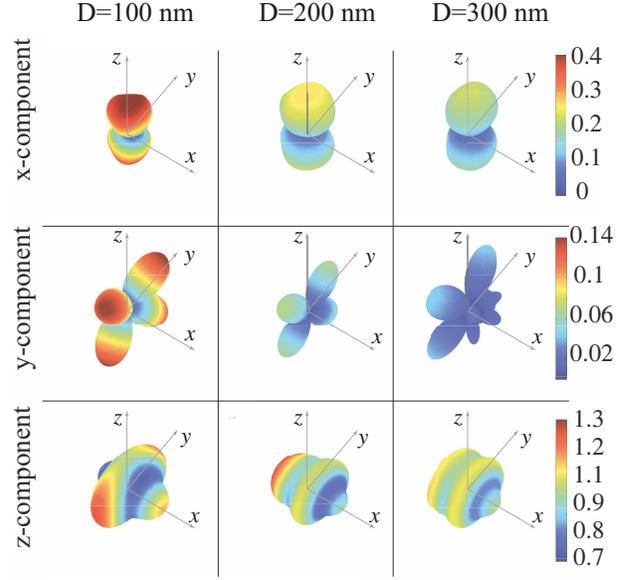}
\caption{The three-dimensional angular dependencies of the relative increase of the components $l=x,y,z$ of the smaller nanoparticle magnetic moment ($|m_2^{(l)}/\alpha^M_2 H_0|$) for the different distances $D$=100, 200, and 300 nm.}
\label{fig_promezh}
\end{figure}
\begin{figure*}[!t]
\includegraphics[width=0.99\textwidth]{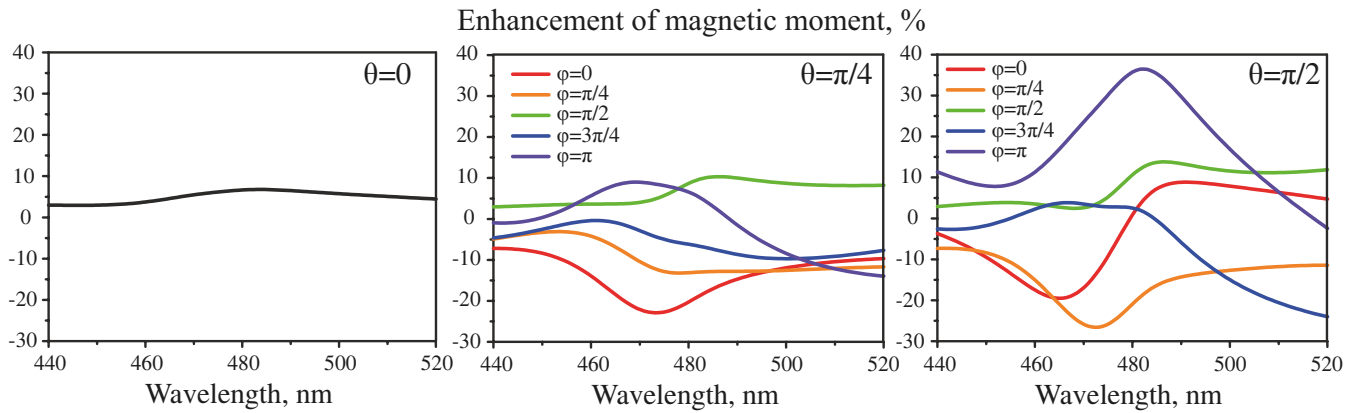}
\caption{The spectral dependencies of the enhancement of magnetic polarizability [$\left(|{m_2}/{\alpha^M_2 H_0}|-1\right)\cdot 100\%$] of the smaller nanoparticle for $D=168$~nm and for different angles $(\varphi,\theta)$ on wavelength. The data for $\theta \in$ ($\pi /2$, $\pi$) and $\varphi \in$ ($\pi $, $2\pi$) repeats the presented results, therefore they are omitted. For the case of $\theta=0$ the lines for all $\varphi$ are coincide.}
\label{fig_05:spectr_results}
\end{figure*}

\section{Vectorial structure of magnetic moments}

In the subsequent studies the illumination was chosen to propagate along $x$-axis and being polarized along $y$. There are three geometrical parameters, affecting the magnetic polarizability of the smaller particle: the distance between the spheres' centres $r_{21}=R_1+R_2+D$, and the angles $\theta$ and $\varphi$. The angular dependence of the induced magnetic moment components of the smaller particle for 3 characteristic separation distances $D=100$, $200$, and $300$ nm (gap between the nanoparticles' surfaces) will be studied next. In the case of noninteracting nanoparticles there is the only one non-zero component of nanoparticle's magnetic moment $m^{(z)}_2$ -- codirected with the magnetic field. In the case of bianisotropic coupling there are also $x$- and $y$- non-zero components of the magnetic moment ($m^{(x)}_2$, $m^{(y)}_2$). Fig.~\ref{fig_promezh} shows the three-dimensional angular dependencies of normalized magnetic moment components ($|m_2^{(l)}/\alpha^M_2 H_0|$, $l=x,y,z$) of the smaller nanoparticle, where $\alpha^M_2$ is the polarizability of the single smaller nanoparticle. These results have been obtained by exact solving Eqs.~(\ref{eg:all_p_m}) taking into account the electric and magnetic responses of both nanoparticles. As it could be seen, the induced components are smaller than the main moment. Those additional vectorial components of magnetic moments increase when the distance between the particles goes down and may being up to 50\% for $D<50$ nm. But it should be noted that for such small distances between the particles accuracy of CEMDA method is low, sow this value is rough~\cite{evlyukhin-optical-2010}.  When the distance between the particles goes up, secondary vectorial components of magnetic moment decrease. Therefore, the overall variation of magnetic moments is mainly determined by its $z$-component. Diagrams of $z$-component of the magnetic moment are asymmetrical and their forms are dissimilar for different separation distances between the particles ($D$). For $D\geq300$~nm enhancement of magnetic moment is weak, because of the vanishing coupling between the particles. For $D>$2 $\mu$m this diagram is symmetrical with good accuracy, as it replicates the performance of the isolated particle. For smaller distances $D$ the maximum of magnetic moment corresponds to the cases ($\theta=\pi/2$, $\varphi=\pi/2$ and $\varphi=3\pi/2$), and for bigger distances maximum of magnetic moment corresponds to the case ($\theta=\pi/2$, $\varphi=\pi$). It should be noted that case ($\theta=\pi/2$, $\varphi=\pi/2$ and $\varphi=3\pi/2$) corresponds to magnetic dipoles coupling only, while the case ($\theta=\pi/2$, $\varphi=\pi$) corresponds to electric-magnetic dipoles coupling only. Additional analysis shows that the maximum of relative magnetic moment enhancement is nearly $36\%$ and it is achieved for $D=168$~nm for the case of ($\theta=\pi/2$, $\varphi=\pi$). The performance of the dimer with this separation distance will be investigated in details hereafter. Both induced magnetic and electric moments of the smaller nanoparticle were calculated using the theoretical model CEMDA. Enhancement of the magnetic polarizability of smaller nanoparticle $\left(|m_2/{\alpha^M_2 H_0}|-1\right)\cdot 100\%$ for different angular arrangements (the distance $D$=168 nm is kept constant) is shown in the Fig.~\ref{fig_05:spectr_results}. These dependences show that the effect has a resonance character. Moreover, the maximum value is achieved at the wavelength of 480 nm i.e. at the electric dipole resonance of bigger nanoparticle and magnetic dipole resonance of smaller one underlining the impact of the resonant nature.

\section{Outlook and conclusions}

Coupled particles approach for controlling magnetic polarizabilities of nanoscale spheres was proposed. While standalone nanoparticles allow obtaining dipolar and high-multipolar resonant responses at the desired wavelength in the visible range specifying their radii and materials, a system of two coupled nanoparticles possesses more degrees of freedom. Altering the radii of both nanoparticles allows to investigate the impact of all the combinations of multipolar coupling effects on the properties of the system. Naturally, the amplitude of the effects is dependent on their mutual displacement, and it decreases at longer distances between the particles, as it was proved both analytically and numerically in the case of dipolar magnetic-electric coupling. Symmetry considerations allow exciting only one dominant induced electric and magnetic moment component when a plane wave is incident upon a single nanoparticle. It was shown, that for a nanoparticle system, altering the spherical angles and distance between nanoparticles allows to excite all vectorial components of moments simultaneously. The secondary components being up to 50$\%$ of the amplitude value of dominant ones. Furthermore, these parameters also define the spectral position of deeps and peaks of the dominant electric moment components and amplitude values, allowing full on-demand control of the electromagnetic properties of the system of coupled nanoparticles.  The proposed approach can find use in designs of more complex structures such as all-dielectric metamaterials
and metasurfaces with strong magnetic responses.

\section{Acknowledgments}

This work was financially supported by the Ministry of Education and Science of the Russian Federation ($\#$14.584.21.0009 10).


%

\end{document}